\documentclass[sigconf]{aamas} 
\usepackage{balance} % for balancing columns on the final page

\setcopyright{ifaamas}
\acmConference[AAMAS '23]{Proc.\@ of the 22nd International Conference
on Autonomous Agents and Multiagent Systems (AAMAS 2023)}{May 29 -- June 2, 2023}
{London, United Kingdom}{A.~Ricci, W.~Yeoh, N.~Agmon, B.~An (eds.)}
\copyrightyear{2023}
\acmYear{2023}
\acmDOI{}
\acmPrice{}
\acmISBN{}

\usepackage{algorithmic}
\usepackage{graphicx}
\usepackage{textcomp}
\usepackage{xcolor}
\usepackage{mathtools}

\usepackage{tablefootnote}
\usepackage{xspace}
\usepackage{adjustbox}
\usepackage{paralist}
\usepackage[colorinlistoftodos]{todonotes}
\title{Developing Multi-Agent Systems with \\ Degrees of Neuro-Symbolic Integration\\{} [A Position Paper]}%A title involving MAS and Neuro-Symbolic Agents [Position Paper]}

\author{Louise Dennis \quad Marie Farrell \quad Michael Fisher\footnotemark\\{} 
 Department of Computer Science, University of
 Manchester, UK}
  \authornote{Farrell and Fisher are both supported by the Royal Academy of Engineering.}
\renewcommand{\shortauthors}{Dennis L. A., Farrell, M., Fisher, M.}

\begin{abstract}
 In this short position paper we highlight our ongoing work on verifiable heterogeneous multi-agent systems and, in particular, the complex (and often non-functional) issues that impact the choice of structure within each agent.
\end{abstract}

\begin{document}

\maketitle
\footnotetext{Corresponding author:\quad \url{michael.fisher@manchester.ac.uk}. }
\section{Introduction}
\begin{figure*}[htbp]
  \includegraphics[width=.37\textwidth]{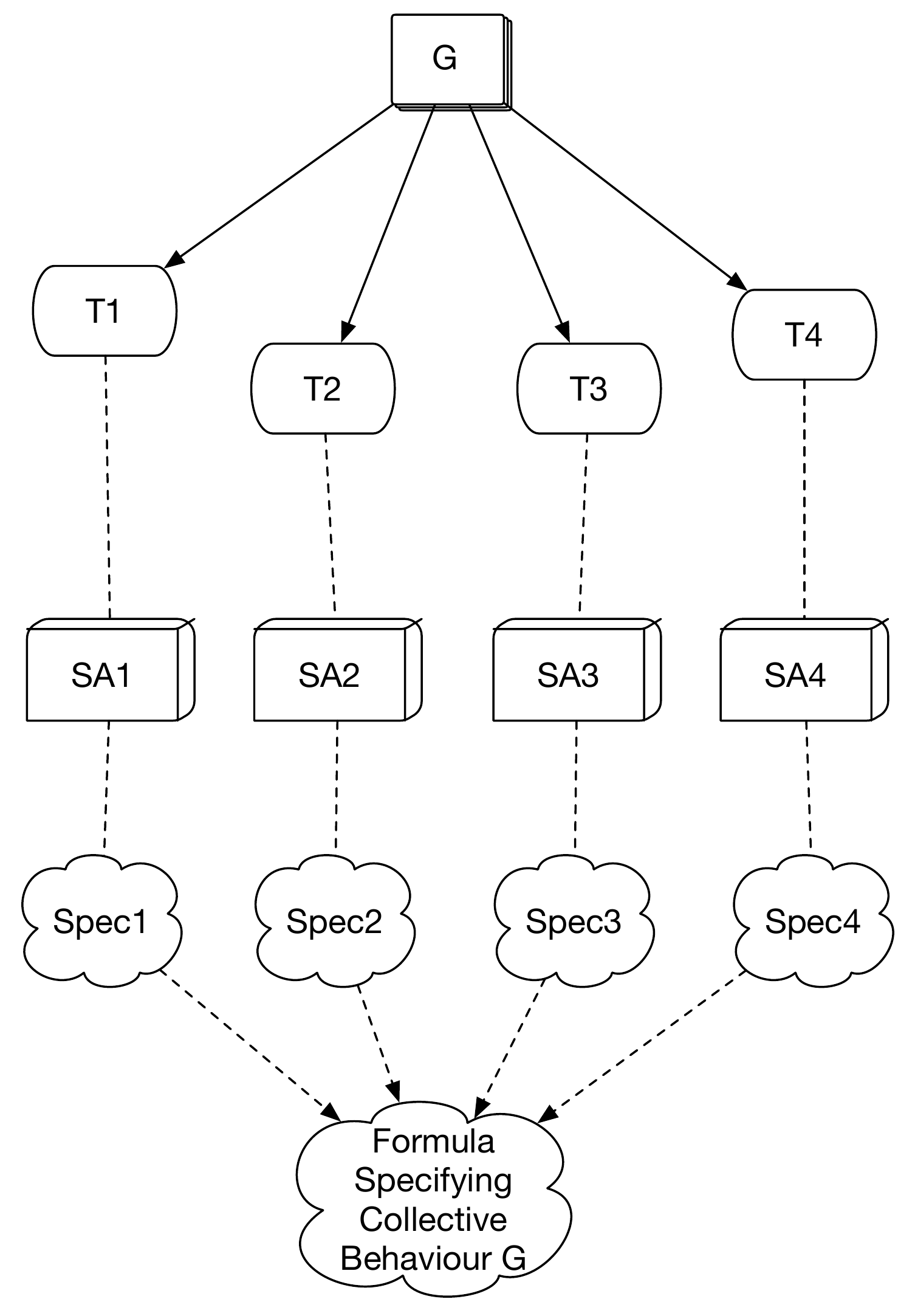}
  \qquad \qquad
  \includegraphics[width=.37\textwidth]{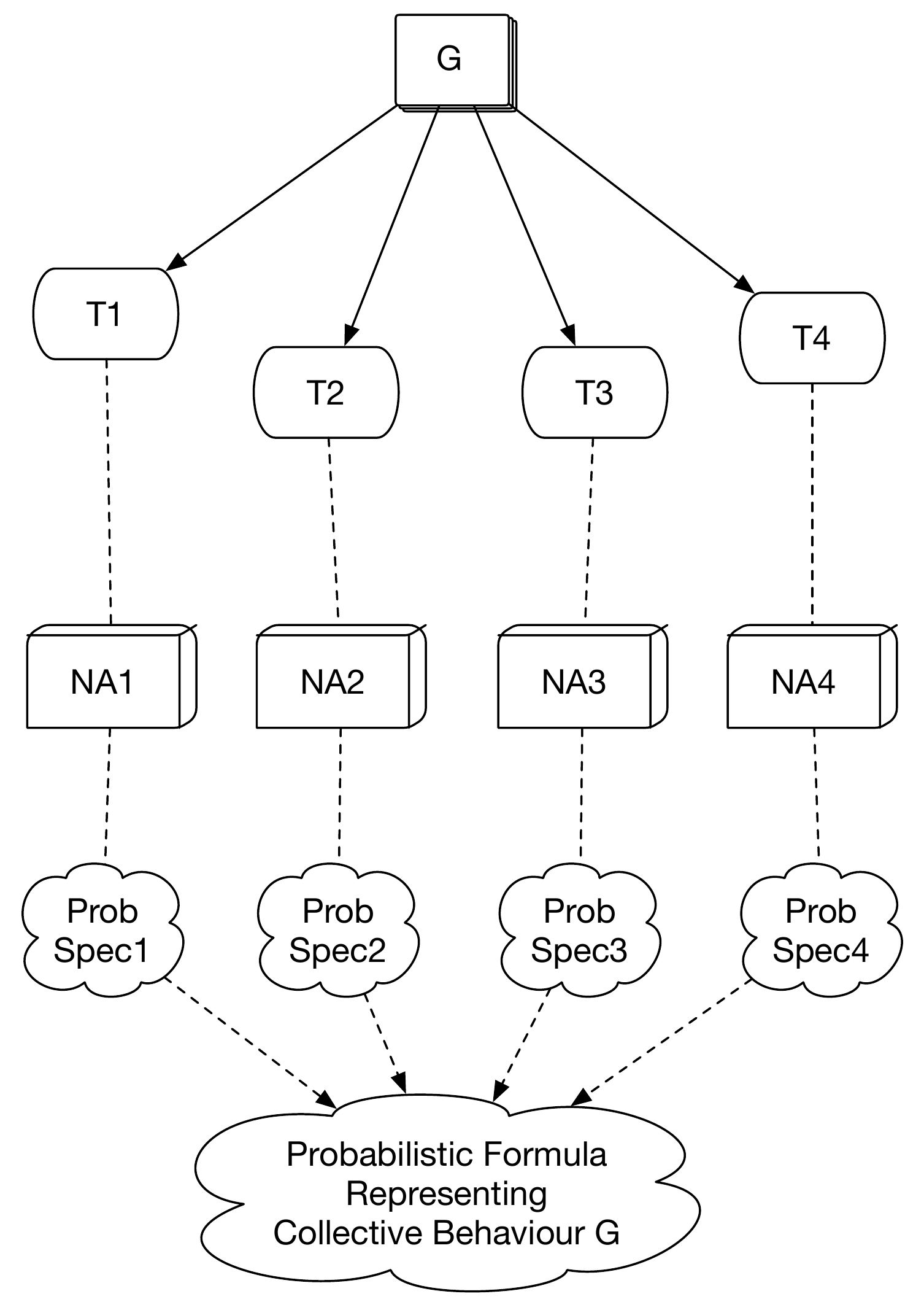}
  \caption{\label{fig1}Decomposing goals into collations of symbolic (left) and neural (right) agents}
\end{figure*}

A traditional way to develop multi-agent systems is to take some
overall goal for the system, say $G$, and decompose this into a set of
tasks (which could be any of actions, updates, sub-goals, etc)~\cite{0033446}. These tasks
are then undertaken by a set of agents, for simplicity one task per
agent. In devising symbolic teams, each agent is symbolic, discharging
its task appropriately (see Figure~\ref{fig1} (left)) giving us a
collection of behaviour specifications (usually in symbolic logic) for
the agents which can then be combined appropriately to provide the
overall description in the form of a symbolic logic formula.

An alternative approach would be to decompose our collective
goal into a set of `neural' agents; see
Figure~\ref{fig1} (right). These agents could be any of (deep) learning
components, adaptive control components, optimisation components,
etc~\cite{ChenLY20}. Again, tasks would be implemented with these agents and the
result would be combined to give an overall outcome. The difference
here is that the behaviour of an individual `neural' component is
described using some stochastic notation, either logical or differential
equations.
\bigskip

\begin{figure*}[htbp]
  \centering
  \includegraphics[width=.37\textwidth]{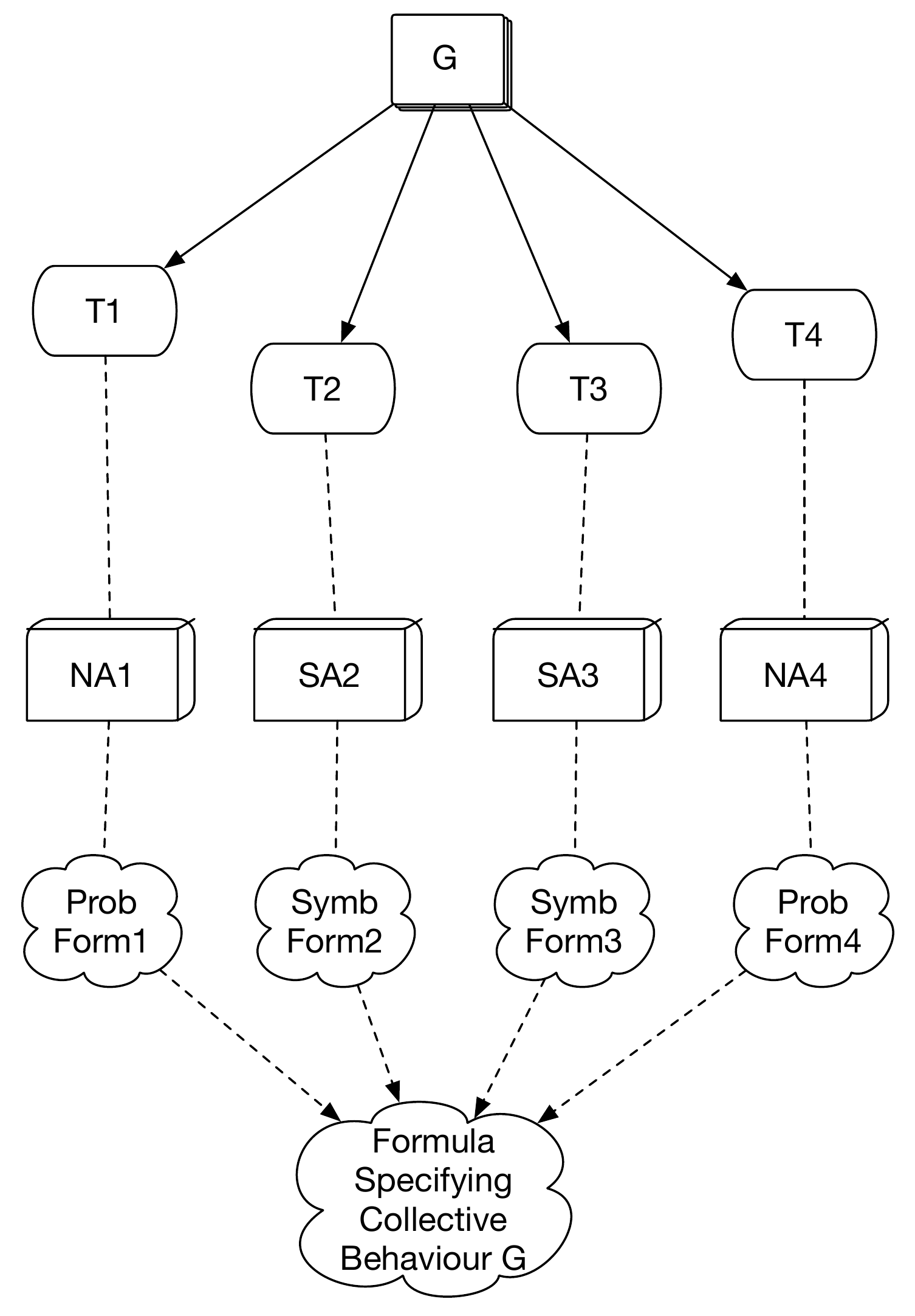}
  \caption{\label{fig2}Decomposing goals into heterogeneous collations of agents}
\end{figure*}

\noindent In reality, however, our multi-agent system is unlikely to be wholly symbolic or wholly neural and so we will get a more heterogeneous multi-agent system such as in Figure~\ref{fig2}.

\section{Position}
The particular issue that we are concerned with here is how, and why, we
choose one particular type of agent over another when constructing our
agent team. Symbolic and `neural' agents generally have quite distinct
properties\footnote{Not unlike Kahneman's `fast' and `slow' categorisations~\cite{Kahneman12}}
\begin{itemize}
\item neural --- fast, able to cope with vast amounts of data/input,
  opaque/stochastic, ...
\item symbolic --- logical, transparent, explainable, verifiable, 
  much slower, may be overwhelmed by data, ...
\end{itemize}
So, our aim is to capture, in the goal specification \textbf{G}, key aspects that need to be considered/achieved relating to this goal. Specifically:
\begin{enumerate}
\item \textsf{Speed} --- Although this might relate to the Multi-Agent System (MAS) as a whole, it is more likely to refer to how quickly a particular task (a decision, recognition, or action) should be completed;

\item \textsf{Transparency} --- Again, while this might refer to the whole MAS, it will more usually relate to specific decisions and actions, exposing exactly what the agents do and why they do it. 
    
    Note: This transparency then forms a basis for both explainability and verifiability~\cite{WinfieldBDEHJMO21}.
    
\item \textsf{Accuracy} --- This refers to decisions that are taken on the basis of potentially inaccurate measurements (e.g. sensor readings, images, etc.). Accuracy-related properties will likely be captured using probabilistic reasoning and/or statistical measures.
\end{enumerate}
Consequently, in specifying the system goal, \textbf{G}, the above aspects must be highlighted. Then, as we decompose this goal into tasks for specific agents, we can assess how well each agent not only can achieve its task but can satisfy the additional speed, transparency, and accuracy requirements. This will surely affect how the MAS is composed and, in many practical cases, we will develop a heterogeneous multi-agent
system as in Figure~\ref{fig2}.

Each agent will be selected to satisfy its speed, transparency, or accuracy requirement and each will have a description of its behaviour. Typically, symbolic agents will have logical descriptions while `neural' agents will have stochastic descriptions (though this may not always be the case).
\begin{itemize}
  \item[] \textbf{Point 1:} we highlight the speed/transparency/accuracy
    requirements in the MAS development process, attempting to ensure that
    the most appropriate style of agent is utilised.
\end{itemize}
\bigskip

\noindent While the above approach can provide us with a description of the distinct agent recommendations, we also need to check that these requirements are actually being achieved. We do this by also
adding \emph{runtime monitors} to check each agent's behaviour and, in particular, whether it is matching the speed/transparency/accuracy requirements. We check speed, in the obvious way by timing progress,
and check transparency by interrogating the agent structures to ensure
that ``what'' is done and ``why'' are explicit.
\begin{itemize}
  \item[] \textbf{Point 2:} since agents may well act sub-optimally,
    we utilise runtime verification~\cite{FalconeHR13} to assess how well each agent is actually matching its speed/transparency/accuracy requirements.
\end{itemize}

\section{Examples}
\paragraph{Autonomous Vehicle.} We consider an autonomous vehicle to actually be a MAS, comprising a wide range of communicating components. We want sensing and recognition to be as fast as possible and so speed is the main criterion here. Although we may veer towards symbolic agents if we require some explanation of recognition, it is much more likely that these aspects will be implemented as some learning component. On the other hand, the decision-making component(s) that handle the high-level decisions that a human driver/pilot/operator used to make will very likely be symbolic agent(s). This facilitates verification, for example against human-level rules, and explanations, for example to authorities or driver who is taking control~\cite{JACIC13} .Lower level actions will undoubtedly be implemented in some adaptive or `neural' control component. 

An interesting element concerns obstacle avoidance. If this is required to be very quick, for example in an emergency situation, then symbolic agents will be too slow. However, if the vehicle has some time to ``think'' then a symbolic agent making a reasoned and explainable decision might be more appropriate. It might well be that both aspects are in the system, with the level of urgency dictating which agent to invoke.

\paragraph{Tele-Operated Robot.} In this situation, speed, accuracy and compliance with the human operator's commands will be the most important aspects. Consequently, very few of the agents need be symbolic. Perhaps the few that are will be concerned with situation awareness (for example, explaining to the distant operator what is happening in the robot's context) or fault diagnosis (for example, explaining what the robot believes has occurred).

\paragraph{Social/Healthcare Teams.} These very complex teams comprise a wide range of agents: some are human (healthcare specialists, etc); some are low-level sensors (detecting movement, temperature, etc); and many are AI-based agents somewhere in between. We could have:
\begin{itemize}
    \item simple cleaning robots, where accuracy (and safety) is central;
    \item health monitoring software, where recognising problems (quickly, accurately) is important but so is explaining to, and interacting with, humans in the environment or team (transparency);
    \item social robots, able to have complex dialogue with humans and able to explain and expand on topics (transparency);
    \item[] and so on. 
\end{itemize} 
In many cases, the strong reliability/verification of the agent, often based on its transparency, will be important.

\paragraph{Cybersecurity:} In this case, a cybersecurity system may use symbolic representations of known attack patterns (based on experience and/or threat analysis) and neural networks to detect previously unknown attacks by analyzing network traffic and system logs. It is especially difficult to distinguish an attack from a software error/failure~\cite{AFT22} so the complex computations performed here by the neural components is beneficial. Here, transparency and speed are particularly important: speed to identify attacks quickly so that mitigations might be invoked and transparency to provide reasoning behind why some behaviours were identified as potential attacks. The latter is useful to build up a knowledge base of why and how previously unidentified attacks occur. A similar approach could be used in other fields such as natural language processing.

\section{Summary}
Decomposing an overall goal into a set of agents that together can
achieve this goal, is well understood. We want to expose additional
constraints, particularly those of speed, accuracy, and transparency, and then
use these in the decomposition/development process to help us select
the variety of agent that is most suitable.

We recognise that, in realistic scenarios, ``things will go wrong''
and so we need to provide dynamic checks that the agents are living up
to their requirements. We do this by inserting a range of runtime
monitors to assess both the speed and transparency of the agents.

The problems related to this involve
\begin{itemize}
\item describing speed, accuracy, and transparency in a flexible but precise way,
\item developing effective, but not intrusive, runtime monitors to assess these, and 
\item extending towards other properties such as flexibility, being able to cope with vast amounts of data, etc.
\end{itemize}
Then, within these complexities we aim to ensure strong verification where appropriate, so supporting the assurance and trustworthiness of complex autonomous teams~\cite{dennis_fisher_2023,BourbouhFMSBD021,FisherMRSWY21}

\bibliographystyle{ACM-Reference-Format} 
%\bibliography{references}

\end{document}